\documentstyle[ltwol]{article}
\newcommand{\rhohat}{{\hat{\rho}}}

\newcommand{\ee}{\varsigma_3}

\newcommand{\btab}{\begin{tabbing}}
\newcommand{\etab}{\end{tabbing}}

\newcommand{\beqn}{\begin{equation}}
\newcommand{\eeqn}{\end{equation}}
\newcommand{\barr}[1]{\begin{array}{#1}}
\newcommand{\earr}{\end{array}}
\newcommand{\beqna}{\begin{eqnarray}}
\newcommand{\eeqna}{\end{eqnarray}}
\newcommand{\btablec}{\begin{table} \begin{center}}
\newcommand{\etablec}{\end{center} \end{table}}

\newcommand{\gapproxeq}{\lower.7ex\hbox{$\;\stackrel{\textstyle>}
{\sim}\;$}}
\newcommand{\plabel}[1]{\label{#1}}

\input{psfig}

\def\be{\begin{equation}}
\def\ee{\end{equation}}
\def\bea{\begin{eqnarray}}
\def\eea{\end{eqnarray}}
\begin{document}
\title{
\begin{flushright} \small{hep-ph/9807433}\\\hspace{1cm}\end{flushright} 
THE 1.4 GEV $J^{PC}=1^{-+}$ STATE AS AN INTERFERENCE OF 
A NON--RESONANT BACKGROUND AND A RESONANCE 
AT 1.6 GEV}

\author{A. DONNACHIE}

\address{Department of Physics and 
Astronomy, University of Manchester, Manchester M13 9PL, UK\\ E-mail: 
ad@a3.ph.man.ac.uk}

\author{P. R. PAGE}

\address{T-5, MS-B283, Los Alamos
National Laboratory, P.O. Box 1663, Los Alamos, NM 87545, USA\\
E-mail: prp@t5.lanl.gov}  


\twocolumn[\maketitle\abstracts{We investigate theoretical interpretations of the 
1.4 GeV $J^{PC}$ exotic resonance reported by the E852 collaboration.
A K--matrix analysis shows that the 1.4 GeV enhancement 
in the E852 $\eta\pi$ data can be understood as an interference of 
a non--resonant Deck--type background and a resonance 
at 1.6 GeV. }]

\footnote{Contributed talk (Abstract 459) presented by P.R. Page 
at the $29^{th}$ Int. Conf. on High Energy Physics (ICHEP'98), Vancouver,
B.C., Canada, July 23--29, 1998.}
Evidence for a $J^{PC}=1^{-+}$ isovector resonance $\rhohat(1405)$ at 1.4 
GeV in the reaction $\pi^{-} p \rightarrow \eta\pi^{-}p$ has
been published recently by the E852 collaboration at 
BNL~\cite{bnletapi}. The mass and width quoted are $1370 \pm 
16^{+50}_{-30}$ MeV 
and $385\pm 40 ^{+65}_{-105}$ respectively.
These conclusions are strengthened by the claim of the Crystal Barrel 
collaboration
that there is evidence for the same resonance in $p\bar{p}$ annihilation 
with a mass of
$1400\pm 20\pm 20$ MeV and a width of $310\pm 50 ^{+50}_{-30}$ MeV 
\cite{cbar}, consistent with E852. However, the Crystal Barrel state 
is not seen as a peak in the $\eta\pi$ mass distribution, but is deduced
from interference in the Dalitz plot. 
Since the $J^{PC}$ of this state is ``exotic'', i.e. it implies 
that it is {\it not}
a conventional meson, considerable excitement has been generated,
particularly because the properties of the state appear to be in conflict 
with
theoretical expectations.

In addition there are two independent indications of a more massive
isovector $J^{PC}=1^{-+}$ exotic resonance $\rhohat(1600)$ in $\pi^{-} N 
\rightarrow
\pi^{+}\pi^{-}\pi^{-} N$.
The E852 collaboration recently reported evidence for a resonance at
$1593\pm 8^{+29}_{-47}$ MeV with a width of $168\pm 20^{+150}_{-12}$ MeV~\cite{bnl97}.
These parameters are consistent with
the preliminary claim by the VES collaboration of a resonance at 
$1.62 \pm 0.02$ GeV with a width of
$0.24\pm 0.05$ GeV~\cite{ves93}. In both cases a partial wave analysis
was performed, and the decay mode $\rho^{0}\pi^{-}$ was observed.
There is also evidence for $\rhohat(1600)$ in $\eta^{'}\pi$ peaking at 
1.6 GeV \cite{ryabchikov97}.
It has been argued that the $\rho\pi,\;
\eta^{'}\pi$ and $\eta\pi$ couplings of this state qualitatively 
support the
hypothesis that it is a hybrid meson, although other interpretations
cannot be entirely eliminated~\cite{page97exo}.

Recent flux--tube and other model estimates~\cite{bcs} and lattice 
gauge 
theory calculations~\cite{perantonis90} for the lightest
$1^{-+}$ hybrid support a mass substantially higher than 1.4 GeV and 
often 
above 1.6 GeV~\cite{page97exo}. Further, on quite general grounds, it 
can be shown that an $\eta\pi$ decay of $1^{-+}$ hybrids is unlikely \cite{page97sel1}. There is 
thus an 
apparent conflict between experimental observation and theoretical 
expectation as far as the 1.4 GeV peak is concerned.

The purpose of the present paper is to propose a resolution of this 
apparent 
conflict. 
We suggest a mechanism whereby an appropriate $\eta\pi$ decay of a
hybrid meson can be generated and argue that there is only one 
$J^{PC}=1^{-+}$ isovector exotic, the lower--mass signal in the E852 
experiment being an artefact of the production dynamics. 
We demonstrate explicitly that is possible
to understand the 1.4 GeV peak observed in $\eta\pi$ as a consequence 
of a 1.6 GeV resonance interfering with a non--resonant Deck--type 
background with an appropriate relative phase.
We do {\it not } propose that there should necessarily be a 
peak at 1.4 GeV; but that if experiment unambiguously confirms
a peak at 1.4 GeV, it can be understood as 
a 1.6 GeV resonance interfering with a non--resonant background.

\section{Interference with a non--resonant background}

The current experimental data on the 1.6 GeV state is consistent with mass 
predictions
and decay calculations for a hybrid meson \cite{page97exo,page97had}. This 
then leaves open the
interpretation of the structure at 1.4 GeV. 

There are two basic problems to be solved. Firstly it is necessary to 
find a mechanism which can generate a suitable $\eta\pi$ width for the
hybrid. Then having established that, it is necessary to provide a 
mechanism to
produce a peak in the cross section which is some way below the real
resonance position.
   
The $\eta\pi$ peak in the E852 data spans the $\rho\pi$ and $b_1\pi$ 
thresholds, so
we propose a Deck--type model~\cite{asc} as a source of a non--resonant 
$\eta\pi$ background. 
We then show that, within the K--matrix formalism, interference
between this background and a resonance at 1.6 GeV can
account for the E852 $\eta\pi$ data. 

\subsection{$\eta\pi$ width of a 1.6 GeV state}

Although the $\eta\pi$ width of a hybrid is suppressed by 
symmetrization selection rules \cite{page97sel1} which
operate on the quark level and have been estimated in QCD sum rules 
to be tiny ($\sim$ 0.3 MeV) \cite{qcdsum},
long distance contributions to this width are possible.

\begin{figure}
\center
\psfig{figure=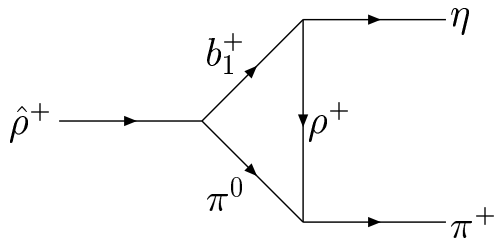,height=1.5in}
\caption{\plabel{fig1} Decay of $\rhohat$ to $\eta\pi$ via final state interactions.}
\end{figure}

An essential ingredient is the
presence of an allowed dominant decay which can couple strongly to the 
channel of interest. In the flux--tube model $b_1\pi$ is such
a dominant decay \cite{page97had}, and it is strongly coupled to $\eta\pi$ by $\rho$ 
exchange (see Figure \ref{fig1}). Diagrams like that in Fig. 1 are 
expected to make the $\eta\pi$ width more appreciable.

\subsection{Non--resonant $\eta\pi$ Deck background} 

The 1.4 GeV peak in the $\eta\pi$ channel occurs in the vicinity of
the $\rho\pi$ and $b_1\pi$ thresholds, and it is therefore natural
to consider these as being responsible in some way for the $\eta\pi$
peak. The Deck mechanism~\cite{asc} is known to produce broad low--mass
enhancements for
a particle pair in three--particle final states, for example in $\pi p 
\rightarrow (\rho\pi) p$. In this latter case, the incident pion 
dissociates into $\rho\pi$, either of which can then scatter off the 
proton~\cite{stod}. At sufficiently high energy and presumed dominance 
of the exchange of vacuum
quantum numbers (pomeron exchange) for this scattering one obtains the
``natural parity change'' sequence $\pi \rightarrow 0^-, 1^+, 2^-....$
(the Gribov--Morrison rule~\cite{gm}). However if the scattering 
involves the exchange of
other quantum numbers then additional spin--parity combinations can be
obtained, including $J^P = 1^-$. This can be seen explicitly in 
ref. \cite{asc} for the reaction $\pi p \rightarrow (\rho\pi) p$
in which the full $\pi p$ scattering amplitude was used, so that
the effect of exchanges other than the pomeron are automatically
included. The $J^P$ sequence
from the ``natural parity change'' dominates due to the dominant 
contribution 
from pomeron exchange, but other spin-parity states are present at a 
non--negligible level. The Reggeised Deck effect can simulate resonances, 
both in terms of the mass distribution and the phase \cite{asc,bow}. 
It can produce 
circles in the Argand plot, the origin of which is the Regge phase factor
exp$[-i{{1}\over{2}}\pi\alpha(t_R)]$.  

It is also important to note that rescattering of the lighter particle 
from the dissociation of the incident beam particle is not a 
prerequisite, and indeed
both can contribute~\cite{stod}. We suggest that in our particular 
case the
relevant processes are (from left to right in Figure \ref{fig2})

\begin{enumerate}

\item $\pi \rightarrow b_1\omega$, $\omega p \rightarrow \pi p$ giving a
$b_1\pi$ final state.
\item $\pi \rightarrow \pi\rho$, $\rho p \rightarrow \eta p$ giving a
$\eta\pi$ final state.
\item $\pi \rightarrow \rho\pi$, $\pi p \rightarrow \pi p$ and $\rho p
\rightarrow \rho p$ giving a $\pi\rho$ final state. 

\end{enumerate}

For each of these processes the rescattering will be predominantly via
$\rho$ (natural parity) 
exchange to give the required parity in the final state.
Obviously process (ii) produces a final $\eta\pi$ state directly, but 
for (i) and (iii) the $b_1\pi$ and $\rho\pi$ final states are required
to rescatter into $\eta\pi$.

The characteristic mass--dependence is a peak just above
the threshold. Thus there are three peaks from our proposed mechanism:
a sharp peak just above the $\eta\pi$ threshold; a broader one at about
1.2 GeV from the $\rho\pi$ channel; and a very broad one at about 1.4
GeV from the $b_1\pi$ channel. The first of these is effectively
removed by experimental cuts, but the net effect of the two latter is 
to produce a broad peak in the $\eta\pi$ channel. Thus invoking this 
mechanism does provide an explanation of the larger width of the $\eta\pi$ 
peak at 1.4 GeV in the E852 data compared to that of the $\rho\pi$ peak at 
1.6 GeV. Because of the resonance--like nature of Deck amplitudes it is also
possible in principle to simulate the phase variation observed. However
as there are Deck amplitudes and the 1.6 GeV resonance, presumably 
produced directly, it is necessary to allow for interference between them. 
We use the K--matrix formalism to calculate this, and also to
demonstrate that the Deck mechanism is essential to produce the 1.4 GeV
peak.

\hspace{3cm}\begin{figure*}
\center
\hspace{-.3cm}\psfig{figure=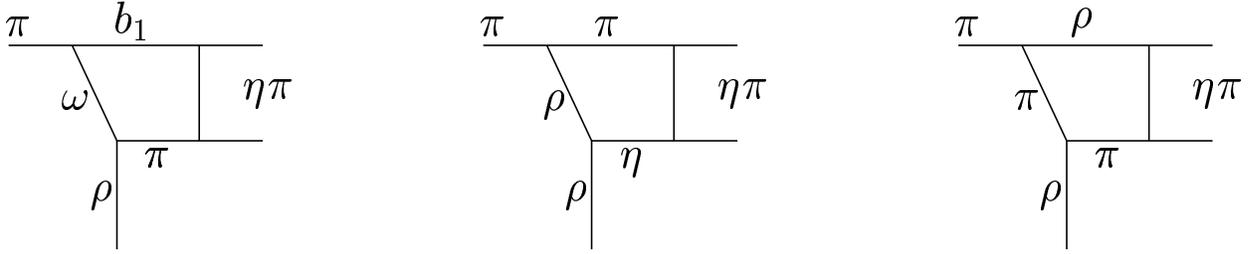,height=1.5in}
\caption{\plabel{fig2} Deck background production in $\eta\pi$.}
\end{figure*}

\subsection{K--matrix with P--vector formalism}

It is straightforward to demonstrate that within the K--matrix 
formalism it is impossible 
to understand the $\eta\pi$ peak at 1.4 GeV as due to a 1.6 GeV state
if only resonant decays to $\eta\pi$, $\rho\pi$ and $b_1\pi$ are 
allowed despite the strong threshold effects in the two latter channels
\footnote{The use of $b_1\pi$ is not critical
here: any channel with a threshold near 1.4 GeV will suffice.}.
We find that for a $b_1\pi$ width of $\approx 200$ MeV and $\eta\pi$ and
$\rho\pi$ widths in the region $1-200$ MeV there is no shift of the peak.
However, when a non--resonant $\eta\pi$ P--wave is introduced, the 
interference between this and the 1.6
GeV state can appear as a 1.4 GeV peak in $\eta\pi$.
 
We have seen that the non--resonant $\eta\pi$
wave can have significant presence at the $b_1\pi$ or $f_1\pi$ threshold 
(called the
``P+S'' threshold), e.g. $1.368$ GeV for $b_1\pi$, because
of the substantial ``width'' generated by the Deck mechanism. Since the 
hybrid is believed to couple 
strongly to ``P+S'' states due to selection rules \cite{page97sel2}, 
the interference effectively shifts the peak in 
$\eta\pi$ down from 1.6 GeV to 1.4 GeV. It is not necessary for the 1.6
GeV resonance to have a strong $\eta\pi$ decay. It is significant that the
E852 experiment
finds $\rhohat$ at $1370\pm 16^{+50}_{-30}$ MeV, near the $b_1\pi$ 
threshold, but not at 1.6 GeV.  
It is possible for a state to peak near the 
threshold of the channel to which it has a strong coupling, assuming
that the (weak) channel in which it is observed has a significant 
non--resonant origin. 

We follow the K--matrix formalism in the P--vector approach as 
outlined in \cite{ait,suhurk}. We assume there to be a $\rhohat$ with
$m_{\rhohat} = 1.6$ GeV  as motivated by the structure 
observed in $\rho\pi$ \cite{bnl97}. The problem is simplified to the 
case where there is decay to two observed channels i.e $\eta\pi$ and
$\rho\pi$, and one unobserved $P + S$ channel. These channels are
denoted 1, 2 and 3 respectively. The production amplitudes and the
amplitude after final--state interactions are grouped together in 
the 3-dimensional P-- and F--vectors respectively. In order to
preserve unitarity \cite{ait} we assume a real and symmetric 
$3 \times 3$ K--matrix. The amplitudes after final--state interactions
and production are related by \cite{ait}  

\beqn  F = (I-iK)^{-1} P \eeqn

We define the widths as

\beqn\plabel{gam} \Gamma_i = \gamma_i^2 \;\Gamma_{\rhohat} \frac{B^2(q_i)}
{B^2(q_i^{\rhohat})}\rho(q_i)\hspace{1cm} i=1,2 \eeqn

\beqn \Gamma_3 = \gamma_3^2 \;\Gamma_{\rhohat}\; \rho(q_3) \eeqn

where $q_i$ is the breakup momentum in channel $i$ from a state of 
effective mass $w$, and $q_i^{\rhohat}$ is the breakup momentum in channel 
$i$ from a state of effective mass $m_{\rhohat}$. The kinematics is 
taken care of by use of the phase space factor

\beqn \rho(q)=\frac{2 q}{w}\eeqn

and the P--wave angular momentum barrier factor

\beqn  B^2(q) = \frac{(q/q_R)^2}{1+(q/q_R)^2}\eeqn

where the range of the interaction is $q_R = 1$ fm $= 0.1973$ GeV.

We assume the experimental width in $\rho\pi$ of $\Gamma_{\rhohat}=168$ MeV \cite{bnl97} to be the total width of the state\footnote{It is found that our results in Fig. 3 are very similar even for a width of 250 MeV.}.
We adopt the flux--tube model of 
Isgur and Paton and
use the $\rho\pi$ and $b_1\pi$ widths which it predicts for a hybrid
of mass 1.6 GeV. Since the model predicts that the branching ratio
of a hybrid to $b_1\pi$ is $59-74$ \% and to $f_1\pi$ is $12-16$ \%
\cite{page97had}, we obtain the $P+S$--wave width to be $120-150$ MeV.  
Analysis of the data
shows that the $\rho\pi$ branching ratio of $\rhohat(1600)$ is $20\pm 2$ \%
\cite{page97exo}, corresponding to a $\rho\pi$ width of 
$30-37$ MeV.  This is consistent with flux--tube model predictions
of $9-22$ \% \cite{page97had}. For the simulation we use a $b_1\pi$ width
of 120 MeV, a $\rho\pi$ width of 34 MeV, and an $\eta\pi$ width of 14 MeV,
well within the limits set by the doorway calculation. We neglect other 
predicted modes of
decay since we restrict our analysis to three channels.

The K--matrix elements are 

\beqn \plabel{kmat} K_{ij} = \frac{m_{\rhohat}\sqrt{\Gamma_i \Gamma_j}}
{m_{\rhohat}^2-w^2} + c_{ij}\eeqn

where $c_{ij}$ includes the possibility of an unknown background.

In the simulation we assume that the Deck terms can be treated as 
conventional resonances.
This is not necessary, but is done to reduce the number of free parameters. 
We assume that the $\eta\pi$ Deck amplitude is produced predominantly via 
the $b_1\pi$ and $\rho\pi$ channels, and so is modelled as a resonance at a mass $m_{b1} =
1.32$ GeV and a width $\Gamma_{b1} = 300$ MeV. This width fits
the E852 data at low $\eta\pi$ invariant masses (see Figure \ref{3ch}a).
The $\rho\pi$ background is assumed
to peak at a mass $m_{b2} = 1.23$ GeV with a width $\Gamma_{b2} = 400$ MeV, 
which when plotted as an invariant mass distribution effectively peaks
at $\sim 1.15$ GeV, in agreement with detailed Deck calculations 
in the $1^{++}$ wave \cite{asc}.

We incorporate the $\eta\pi$ and $\rho\pi$ Deck background by putting 
$c_{ij}=0$ except for

\beqn c_{11}=\frac{m_{b1}\Gamma_{b1}}{m_{b1}^2-w^2} \hspace{.8cm}
c_{22}=\frac{m_{b2}\Gamma_{b2}}{m_{b2}^2-w^2}
\eeqn

The widths 
are defined analogously to Eq. \ref{gam} as

\beqn \Gamma_{bi} = \gamma_{bi}^2 \;\Gamma_{\rhohat}\; \frac{B^2(q_i)}
{B^2(q_i^{b})}\rho(q_i)\hspace{1cm} i=1,2  \eeqn

where $q_i^b$ is the breakup momentum from a state of effective mass 
$m_{bi}$ (for $i=1,2$).

The production amplitudes are given by

\beqn P_i = \frac{m_{\rhohat} V_{\rhohat} \sqrt{\Gamma_i 
\Gamma_{\rhohat}}}{m_{\rhohat}^2-w^2} + c_i \eeqn

where the (dimensionless) complex number $V_\rhohat$ measures the 
strength of the production of $\rhohat$. We take $c_3=0$ and 

\beqn
c_1 = \frac{m_{b1} V_{b1} \sqrt{\Gamma_{b1} 
\Gamma_{\rhohat}}}{m_{b1}^2-w^2} \hspace{.6cm}
c_2 = \frac{m_{b2} V_{b2} \sqrt{\Gamma_{b2} 
\Gamma_{\rhohat}}}{m_{b2}^2-w^2} 
\eeqn
where the complex numbers $V_{bi}$ gives the production strengths of the 
Deck background in channel $i$.  

The results of this fit are shown in Fig. \ref{3ch} and clearly provide a 
good
description of the $\eta\pi$  data \cite{bnletapi,suhurk}. 

We briefly discuss the results.
Fig. \ref{3ch}a indicates a steep rise for low invariant $\eta\pi$ masses, 
and a slow fall for 
large $\eta\pi$ masses. This naturally occurs because of the presence of 
the resonance at 1.6 GeV in the high mass region, which shows as a 
shoulder in our fit. 
Figure \ref{3ch}b reproduces the experimental slope and phase change
in $\eta\pi$ \cite{suhurk}. One might find this unsurprising, since the
background changes phase like a resonance. However, we have confirmed,
by assuming a background that has constant phase as a function of
$\eta\pi$ invariant mass, that the experimental phase shift is still
reproduced. The experimental phase shift is hence induced by the 
resonance at 1.6 GeV.

Without the inclusion of a dominant $P+S$--wave channel the $\eta\pi$ 
event shape
clearly shows two peaks, one at 1.3 GeV and one at 1.6 GeV, which is not 
consistent
with the data \cite{bnletapi}. The phase motion is also more pronounced in 
the 
region between the two peaks than that suggested by the data \cite{suhurk}.
The r\^{o}le of the dominant $P+S$--channel is thus that at invariant 
masses between
the two peaks, the formalism allows coupling of the strong $P+S$ channel 
to $\eta\pi$, so that the
$\eta\pi$ appears stronger than it would otherwise, interpolating between 
the peaks at
1.3 and 1.6 GeV, consistent with the data \cite{bnletapi}. A dominant $P+S$
 decay of the $\rhohat$
is hence suggested by the data. 

\begin{figure}
\center
\vspace{-2cm}
\psfig{figure=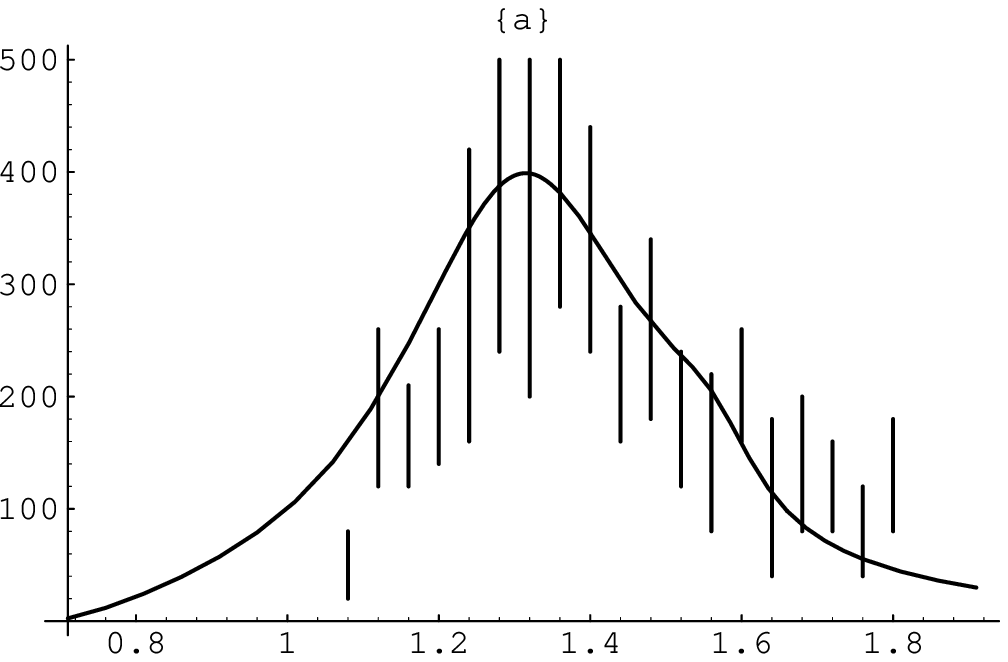,height=3.3in}
\vspace{-1.5cm}
\psfig{figure=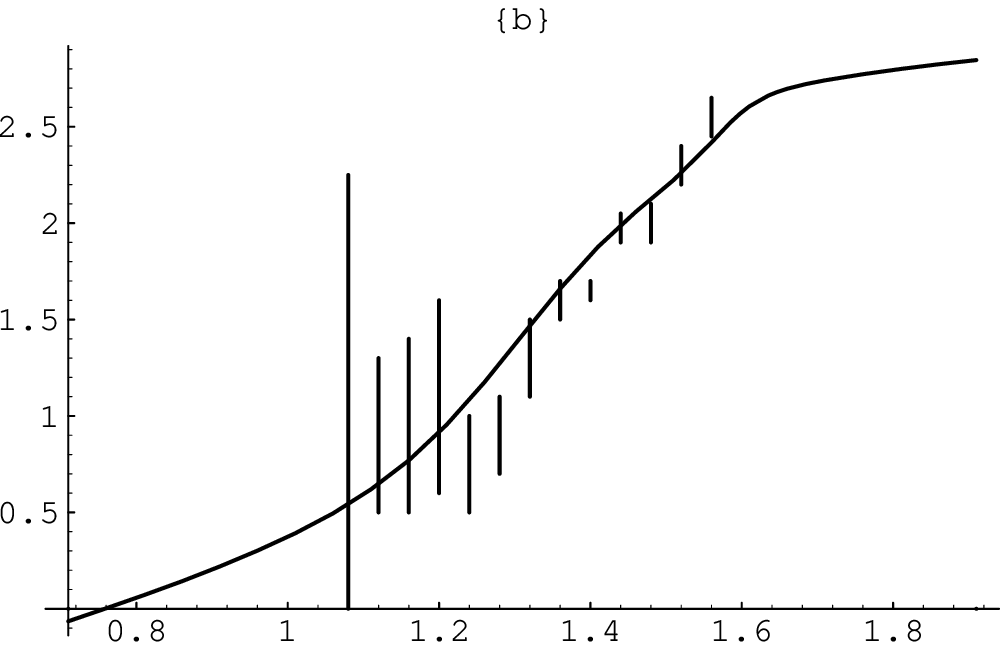,height=3.3in}
\caption{\plabel{3ch} Results of the K--matrix analysis.  (a) The events 
($|F_1|^2$) in $\eta\pi$ as compared to experiment \protect\cite{bnletapi};
(b) The phase (of $F_1$) in $\eta\pi$ compared to experiment 
\protect\cite{suhurk}.
%; (c) The events  ($|F_2|^2$) in 
%$\rho\pi$; (d)  The phase (of $F_2$) in $\rho\pi$. 
The invariant mass $w$ 
is plotted on the horisontal axis in GeV. When the phase is plotted it is 
in radians, with the overall phase {\it ad hoc}. The parameters of the simulation are $m_{\rhohat}=1.6$ GeV, $\Gamma_{\rhohat}=168$ MeV \protect\cite{bnl97}, $\gamma_1 = 0.31$, $\gamma_2 = 0.52$, $\gamma_3 = 1.49$, $m_{b1} = 1.32$ GeV, $m_{b2} = 1.23$ GeV, $\gamma_{b1} = 1.53$, $\gamma_{b2} = 2.02$, $V_{b1}/V_{\rhohat} = 2.05 e^{2.77 i}$, $V_{b2}/V_{b1} = 0.35 e^{1.6 i}$. $V_{\rhohat}$ sets the overall magnitude and phase, which is not shown. None of the ratios of production strengths should be regarded as physically significant, since the K--matrix formalism allows for the introduction of additional parameters in the modelling of the backgrounds, which would change the values of these ratios. The plots shown here are only weakly dependent
on the $\rho\pi$ parameters $\gamma_{b2}$ and $V_{b2}$.
The parameters have been chosen to fit both the
$\eta\pi$ data \protect\cite{bnletapi} and the preliminary $\rho\pi$ data
 \protect\cite{bnl97}. 
Experiment
has not been able to eliminate the possibility that the low mass peak 
in $\rho\pi$ is 
due to leakage from the $a_1$.
The background
amplitude in $\rho\pi$ is being used as a means of parametrising all forms of background into the $\rho\pi$ channel,
including leakage or Deck. }
\end{figure}

\section{Discussion}

We have argued that on the basis of our current understanding of meson
masses it is implausible to interpret the 1.4 GeV peak seen in the
$J^{PC} = 1^{-+}$ $\eta\pi$ channel by the BNL E852 experiment as 
evidence for an exotic resonance at that mass. We acknowledge that
this is not a proof of non--existence and note the Crystal Barrel
claim for the presence of a similar state at $1400 \pm 20\pm 20$ MeV in
the reaction $p\bar p \rightarrow \eta\pi^+\pi^-$. However this is
not seen as a peak and is inferred from the interference pattern
on the Dalitz plot. It has not been observed in other channels in
$p\bar p$ annihilation at this mass, which is required for
confirmation. So at present we believe that the balance of 
probability is that the structure does not reflect a real resonance.

Given this view, it is then necessary to explain the data and in
particular the clear peak and phase variation seen by the E852 
experiment. Additionally the observation of the peak only in the
$\eta\pi$ channel, which is severely suppressed 
by symmetrization selection rules, requires justification. We have dealt with
these two questions in reverse order. We first suggest
final--state interactions can generate a sizable $\eta\pi$ decay.
We then suggest that the E852 $\eta\pi$ peak is due
to the interference of a Deck--type background with a hybrid
resonance of higher mass, for which the $\rhohat$ at 1.6 GeV
is an obvious candidate. This mechanism also provides the
natural parity exchange for the former which is observed experimentally.
The parametrization of the Deck background is found not to be critical.

A key feature in our scenario is the presence of the large ``$P+S$''
amplitude which drives the mechanism. This should be observable both
as a decay of the 1.6 GeV state and as a lower--mass enhancement due to
the Deck mechanism. Depending on the relative strength of these two 
terms the resulting mass distribution could be considerably distorted
from a conventional Breit--Wigner shape as the Deck peak is broad and
the interference could be appreciably greater than in the $\rho\pi$ channel.

\end{document}